\documentclass[%
reprint,
superscriptaddress,
 amsmath,amssymb,
 prl,
]{revtex4-2}

\usepackage[normalem]{ulem}
\usepackage{graphicx}
\usepackage{dcolumn}
\usepackage{bm}
\usepackage{hyperref}

\usepackage[utf8]{inputenc}
\usepackage[T1]{fontenc}
\usepackage{mathptmx}
\usepackage{etoolbox}
\usepackage{verbatim}

\usepackage{caption}
\usepackage{subcaption}
\usepackage{amsmath}
\usepackage{amssymb}
\usepackage{xcolor}
\usepackage{hyperref}

\begin{document}



\title{Quasi-static transverse electric field driven electron acceleration in relativistic laser matter interaction}

\author{Ameya Parab}
\affiliation{Tata Institute of Fundamental Research, 1 Homi Bhabha Road, Colaba, Mumbai 400 005, India.}

\author{Bhooshan Paradkar}%
\affiliation{UM-DAE Centre for Excellence in Basic Sciences, University of Mumbai, India.}

\author{Aparajit C.}
\affiliation{Tata Institute of Fundamental Research, 1 Homi Bhabha Road, Colaba, Mumbai 400 005, India.}

\author{Anandam}
\affiliation{Tata Institute of Fundamental Research, 1 Homi Bhabha Road, Colaba, Mumbai 400 005, India.}

\author{Sk Rakeeb}
\affiliation{Tata Institute of Fundamental Research, 1 Homi Bhabha Road, Colaba, Mumbai 400 005, India.}

\author{Sagar Dam}
\affiliation{Tata Institute of Fundamental Research, 1 Homi Bhabha Road, Colaba, Mumbai 400 005, India.}

\author{Prashant Kumar Singh}
\affiliation{Tata Institute of Fundamental Research, Gopanapally, Serilingampally, Telangana 500046, India.}

\date{\today}

\begin{abstract}

Achieving significant energy gain in laser-driven relativistic electron beams remains challenging due to dephasing between the accelerating laser field and the electrons. We show that transverse electric fields, when aligned with the plane of laser polarization, can mitigate dephasing and enable substantial energy gain without compromising beam directionality. As a practical realization, we propose a two-laser scheme in which one laser generates the transverse field while the other drives electron acceleration. By tailoring the interaction geometry, this configuration sustains phase locking, enhances energy transfer, and opens a pathway toward compact, high-efficiency electron accelerators.

\end{abstract}
\maketitle

{ Acceleration of electrons in the presence of intense electromagnetic radiation is a fundamental problem in plasma physics, with far-reaching implications for the development of compact, laser-driven tabletop accelerators\cite{Optica_rocca,Wilks1993,YPing_abs,Malka2008}. Conventionally, the transfer of energy from the laser to electrons is mediated by quasi-static longitudinal electric field, as examplified in laser wakefield electron acceleration~\cite{tajima1979laser,esarey2009physics}. On the other hand, direct transfer of energy from laser electric field to an electron requires breaking of symmetry constrains set by the Woodword-Lawson theorem ~\cite{Lawson,woodwork}. The strategies proposed to achieve such energy transfer include use of longitudinal electric field associated with a Gaussian laser beam \cite{VLA_semiinfinite}, excitation of surface plasmons through corrugated targets~\cite{Macchi,Riconda_PoP2015_SPW_electron_scaling},external magentic field\cite{ext_mag_1,ext_mag_2,ext_mag_3_axial}, phase-locking with longitudinal electric field \cite{ext_long_1}, transverse oscillations in underdense plasma channel \cite{ext_long_2} and so-called direct laser acceleration(DLA)/vaccum laser acceleration(VLA) mechanisms \cite{VLA_OG,VLA_natphy,VLAsingh2022vacuum}.\\}
{ The key challenge in direct energy transfer from the laser lies in keeping an electron `locked' with the correct phase of laser electric field to extract net energy gain during the interaction. This critically depends upon the phase slippage between the laser and the electron, which in turn is set by the relative speed between them. Therefore, one can envisage that a longitudinal electric field, depending upon its direction, will push the electron to reduce its slippage with the accelerating phase of the laser electric field. This idea was exploited to explain existence of super-ponderomotive electrons in the laser-plasma interaction \cite{ext_long_1,arefiev_paramatric_amplification}. However, the role of the transverse quasi static electric field on phase locking conditions was not fully understood. In this letter, we address this issue and demonstrate that phase locking is extremely sensitive to the presence of transverse electric field, significantly affecting the electron dynamics leading to direct transverse of energy from the laser field. 

As a proof of concept, we propose a laser-plasma interaction experiment to demonstrate efficient acceleration of electrons in presence of transverse quasi-static electric field. Through detailed analysis of electron trajectories, obtained from Particle-In-Cell (PIC) simulations, we unequivocally demonstrate the `phase-locking' of electrons with the laser field due to the presence of transverse electric field.\\}
{ We begin with analyzing 1D3V motion of a single electron in presence of an electromagnetic wave and quasi-static electric field. The motion is described in units such that time ($t$) and position ($\mathbf{x}$) are normalized by laser frequency $\omega_0$ and wavenumber $k_0$ i.e. $t \rightarrow \omega_0t \, , \mathbf{x} \rightarrow k_0 \mathbf{x}$. Accordingly, the velocity and momentum are normalized as $\mathbf{v} \rightarrow \mathbf{v}/c \, , \mathbf{P} \rightarrow \mathbf{P}/m_ec$. The electric and magnetic fields are normalized as $\mathbf{E} \rightarrow \dfrac{e\mathbf{E}}{m_e\omega_0c} \, , \mathbf{B} \rightarrow \dfrac{e\mathbf{B}}{m_e\omega_0c}$. Final the fields associated with the laser, propagating along $x$-direction, are described by the vector potential $\mathbf{A} \rightarrow \dfrac{e\mathbf{A}}{m_ec^2}$. Taking laser vector potential of the form $\mathbf{A} = A_y(y,\tau) \hat{y}$, with $\tau = t - x$, the equation of motion of a single electron in presence of the laser and quasi-static longitudinal ($E_x^{0}$) and transverse ($E_y^{0}$) electric fields can be described as
\begin{subequations}\label{eq:EOM}
\begin{align}
\frac{d P_x}{dt} &
           = -E_x^{0}+\, v_y \frac{\partial A_y}{\partial \tau}\,,
\label{eq:EOMx}\\[3pt]
\frac{d P_y}{dt} &
           = -E_y^{0}+\, (1 -v_x) \frac{\partial A_y}{\partial \tau}\,,
\label{eq:EOMy}\\[3pt]
\frac{d P_z}{dt} &
           = 0,
\label{eq:EOMz}\\[3pt]
\frac{d \gamma}{dt} &= -v_x E_x^{0} \,- v_y\left(E_y^{0} \,- \frac{\partial A_y}{\partial \tau}\right) \,,
\label{eq:EOMgamma}
\end{align}
\end{subequations}
where $\gamma = \sqrt{1 + P_x^2 + P_y^2 + P_z^2}$. In case of plane waves, i.e. $A_y = A_y(\tau)$, we get following equations:
\begin{subequations}\label{eq:EOM_1}
\begin{align}
\frac{d}{dt}\left(P_y - A_y\right) &
                  =  -E_y^{0}\,, 
\label{eq:EOM_perp}\\[3pt]
\frac{d}{dt}\left(\gamma - P_x\right) &
                  =  E_x^{0}(1 - v_x) \,- E_y^{0}v_y\,.
\label{eq:EOM_par}
\end{align} 
\end{subequations}
The phase-slippage of an electron is characterized by $d\tau/dt = 1 - v_x = \left(\gamma - P_x\right)/\gamma$. Integrating Eq.~\ref{eq:EOM_perp} with $p_y(t = 0) = A_y(t = 0)=0$ and introducing $R = \gamma - P_x$ \cite{ext_long_1}, we get the following coupled system of equations to describe the dynamics of the electron.
\begin{subequations}\label{eq:EOM_2}
\begin{align}
    \frac{d \tau}{dt} & 
                      = \frac{R}{R+P_x}\,,
    \label{eq:EOM_tau}\\[3pt]
    \frac{d P_x}{dt} &
                     = -E_x^{0} \,+ \frac{1}{2R}\frac{d A_y^2}{dt} \,- \frac{1}{R} \frac{d A_y}{dt}\int_0^tE_y^{0}dt \,, 
    \label{eq:EOM_px}\\[3pt]
    \frac{d R}{dt}  &
                    = \frac{R}{R+P_x}\left[E_x^{0} \,- \frac{E_y^{0}A_y}{R} \,+ \frac{E_y^{0}}{R}\int_0^tE_y^{0}dt\right] \,.
    \label{eq:EOM_R}
\end{align}
\end{subequations}

\par Numerical solution of Eq.~\ref{eq:EOM_2} for $A_y(\tau) = a_0(1-\exp(-\tau))\cos{\tau}$ is shown in Fig.~\ref{fig:theory}, where the peak value of laser vector potential is taken as $a_0 = 3$. Here, the top panel (a) shows the phase-slippage ($d\tau/dt$), the middle panel (b) shows the instantaneous laser vector potential $A_y$ seen by the electron whereas the bottom panel (c) represents the electron's longitudinal momentum, which is normalized by $a_0^2/2$. In the absence of quasi-static fields ($E_x^{0} = E_y^{0} = 0$), above equations reduce to $R = R_0$, $P_x = P_{x0}\, +\,A_y^2/(2R_0)$ and $d\tau/dt = R_0/(R_0+P_{x0}+A_y^2/(2R_0))$, where $R_0$ and $P_{x0}$ are the initial constants. For an electron starting from rest ($R_0 = 1, P_{x0} = 1$), the longitudinal momentum is expressed as $P_x = A_y^2/2$. This case is shown as a blue dashed line in Fig.~\ref{fig:theory}. From Fig.~\ref{fig:theory}(a) we see that this electron does not get phase-locked with the laser due to oscillatory phase-slipage ($d\tau/dt = 2/(2+A_y^2)$). According the laser field seen by the electron is also oscillatory causing no direct gain from the laser field (Fig.~\ref{fig:theory}(b)). From Fig.~\ref{fig:theory}(c), we see that in the absence of quasi-staic fields, the maximum longitudinal momentum is $a_0^2/2$.  
\par Next we consider the case of pure longitudinal quasi-static electric field i.e. $E_x^0 \neq 0, E_y^0 = 0$. This case with $E_x^0 = -0.2$ and $E_y^0 = 0.0$ is shown by a black dotted line in Fig.~\ref{fig:theory}. Here, the electron becomes super-ponderomotive, consistent with the earlier work on DLA with longitudinal electric fields \cite{ext_long_1}. From Fig.~\ref{fig:theory}(c), we see that the maximum longitudinal momentum is found to be around 10 times that of vacuum acceleration ($a_0^2/2$). In this case, electron experiences a constant push in the laser propagation direction leading to monotonic drop in the phase-slippage (see Fig.~\ref{fig:theory}(a)). The monotonic drop in phase-slippage can be explained from Eq.~\ref{eq:EOM_R}, which can be expressed as $dR/d\tau = E_x^0$, when $E_y^0 = 0$. Therefore, $R$ drops linearly with $\tau$ in the pure longitudinal case. Accordingly, the electron sees slowly varying laser fields as can be seen from black dotted line in Fig.~\ref{fig:theory}(b).  

\par Finally, we discuss the case when both longitudinal ($E_x^0 = -0.2$) and transverse ($E_y^0 = 0.05$) quasi-static fields are present (red solid line in Fig.~\ref{fig:theory}). In this case, we see that the maximum gain in the longitudinal momentum is about 20 times maximum vacuum gain. Note that this is almost twice that of gain obtained with the identical longitudinal electric field (black dotted line). Thus, addition of transverse quasi-static field can substantially influence the direct acceleration of an electron from the laser field. Interestingly, we observe a substantially larger gain in longitudinal momentum when both quasi-static fields are present, despite the increased phase slippage compared to the case with only the longitudinal quasi-static field. This finding suggests that while reduced phase slippage may be a necessary condition for efficient direct laser acceleration (DLA), it is not by itself sufficient; additional mechanisms associated with the combined quasi-static fields play a crucial role in enhancing electron energization. In the present case, the additional positive contribution to gain in $P_x$ comes from the third term in Eq.~\ref{eq:EOM_px}, where $dA_y/dt < 0$ after $t/T_0 > 5$. This term represents additional contribution to $v \times B$ force due to the transverse quasi static electric field. It makes a positive contribution to the longitudinal momentum in every half cycle of the laser, irrespective of the polarity of the transverse electric field.

Additionally, the contribution from the second term also becomes positive after $t/T_0 \simeq 25$, when both $A_y$ and $dA_y/dt$ are negative. In contrast, this term in the pure longitudinal field case contributes significantly less as $A_y \simeq 0$. This explain higher gain with combined quasi-static fields inspite of having higher phase-slippage. In summary, the presence of transverse quasi-static fields can substantially help in achieving the phase locking conditions for direct laser acceleration.    
\begin{figure}[ht!]
    \centering
    \includegraphics[width=\linewidth]{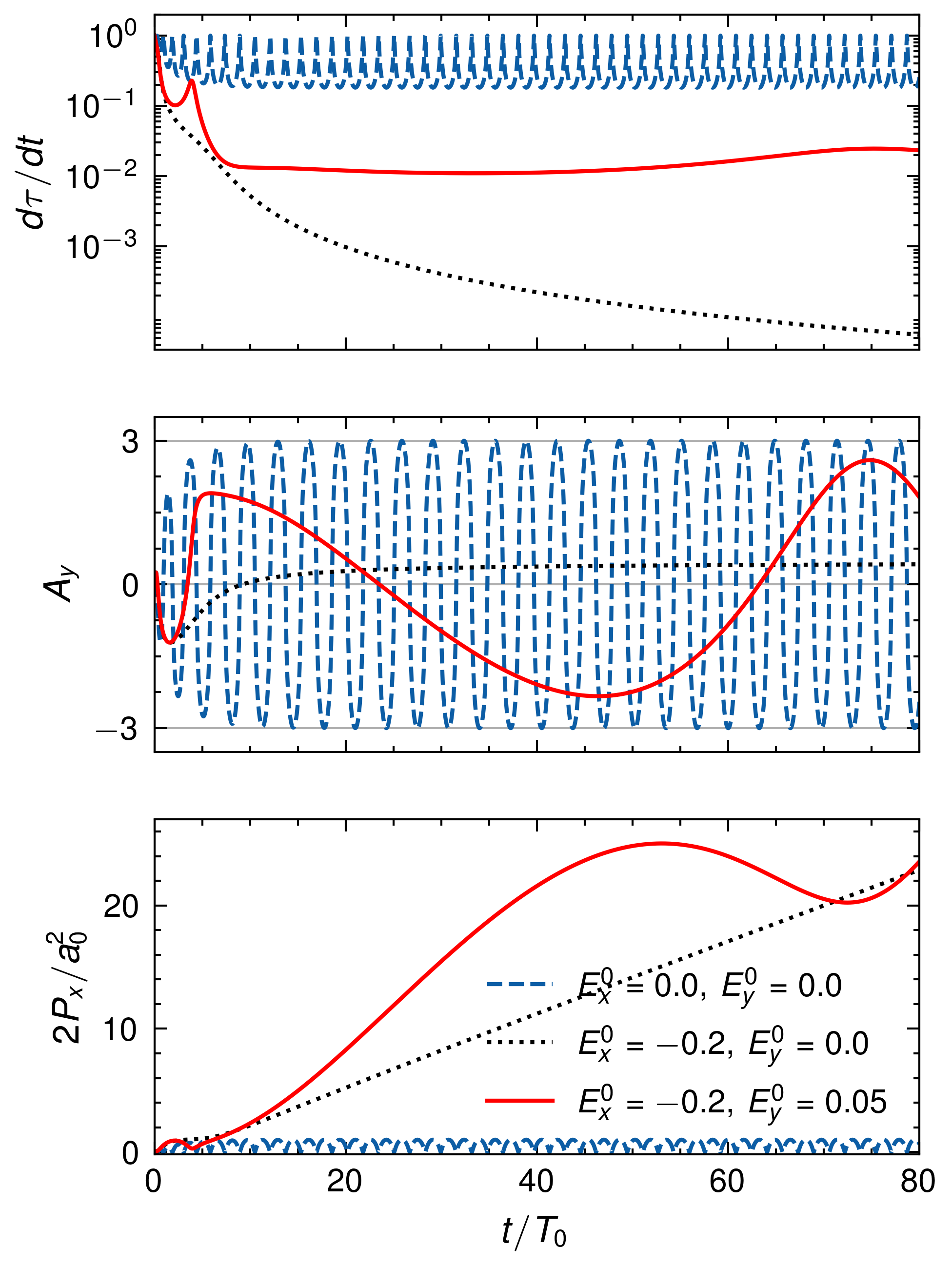}
    \caption{ Numerical solution of Eq.~\ref{eq:EOM_2} with/without quasi-static fields. The electron, initial at rest, moves under the influence of laser vector potential $A_y(\tau) = a_0(1-\exp(-\tau))\cos{\tau}$ with $a_0 = 3$. Fig.(a) shows phase-slippage ($d\tau/dt$); (b) shows the instantaneous laser vector potential $A_y$ seen by the electron and (c) shows the normalized longitudinal momentum.}
    \label{fig:theory}
\end{figure}   
\par Using the physical intuition gained from above discussion, we now propose a realistic laser-plasma experiment to utilize such quasi-static fields for efficient electron acceleration. It should be noted that for such an acceleration to take place, one needs sufficiently large quasi-static fields. In laser-plasma experiments, the fields of such magnitude can be naturally obtained near the sheath formed on the rear side of solid targets. In well-studied Target Normal Sheath Acceleration (TNSA)~\cite{wilks2001energetic,hatchett2000electron,snavely2000intense} mechanism, such strong quasi-static fields are produced due to buildup of a strong positively charged layer at the rear surface to accelerate ions normal to the target. While effective for ion acceleration, this sheath field can simultaneously hinder electron acceleration by decelerating rear-emitted electrons and exacerbating refluxing\cite{nature_refluxing_exp,nilson2008high,buffechoux2010hot}, thereby complicating the efficient generation of high-energy electron beams. Consequently, experimental efforts were focused either on suppressing the sheath field to facilitate electron acceleration~\cite{Two_laser_electron}. Alternatively, we propose that such sheath fields can help in facilitating the DLA, provided another laser is incident on the target rear side for accelerating the electrons.     

\begin{figure}[hb!]
\centering
\includegraphics[width=\linewidth]{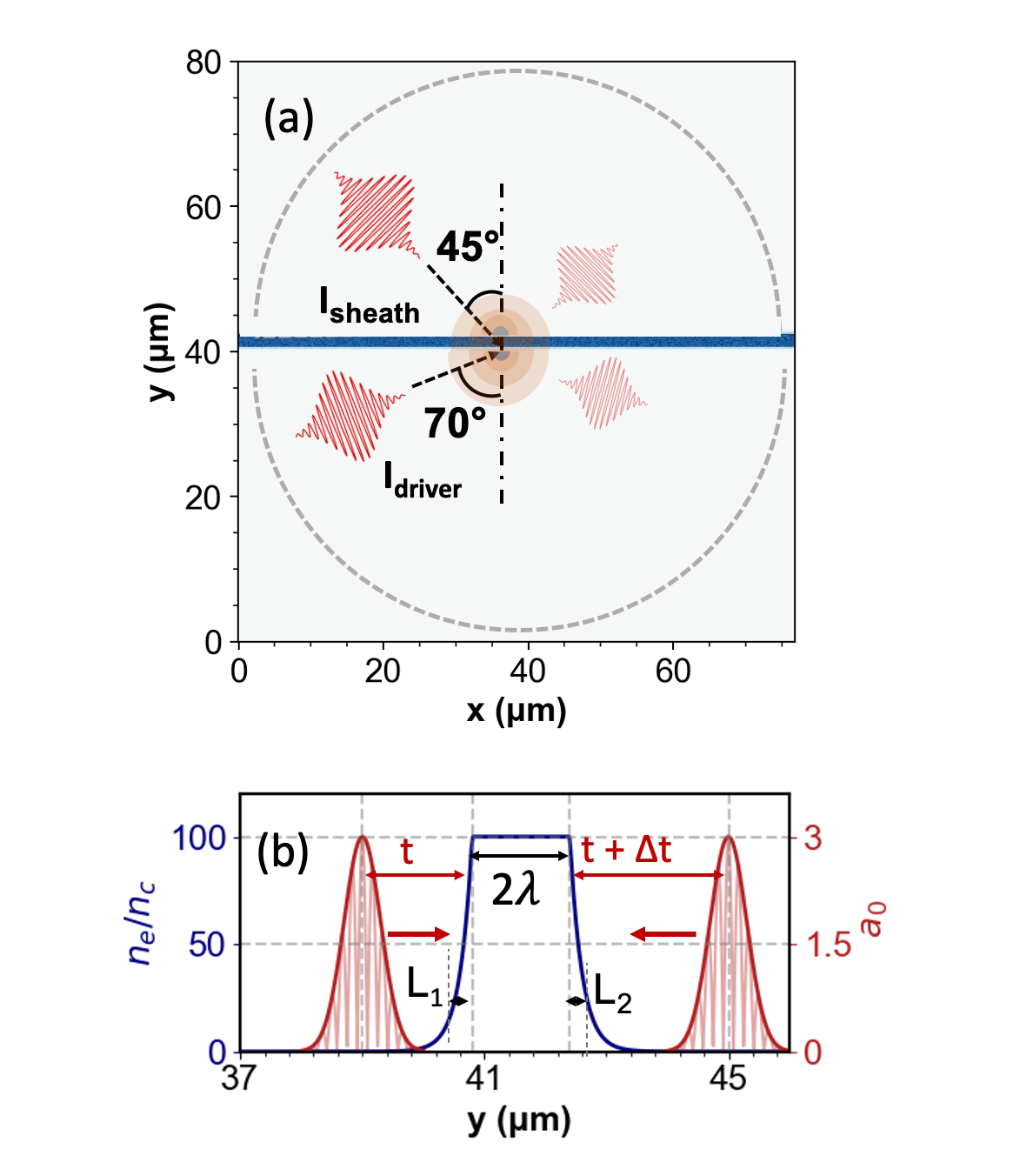}
\caption{1(a) shows 2D PIC Simulation setup with two lasers incident on the either side of the target ($2\lambda$) indicated in blue; 1(b) shows the schematic of lasers incident with a delay of $\Delta t$ on a target with an exponential scale length of $L_1$ and $L_2$.}
\label{fig:setup}
\end{figure}

\begin{figure*}[ht!]
   \centering
   \includegraphics[width=0.9\linewidth]{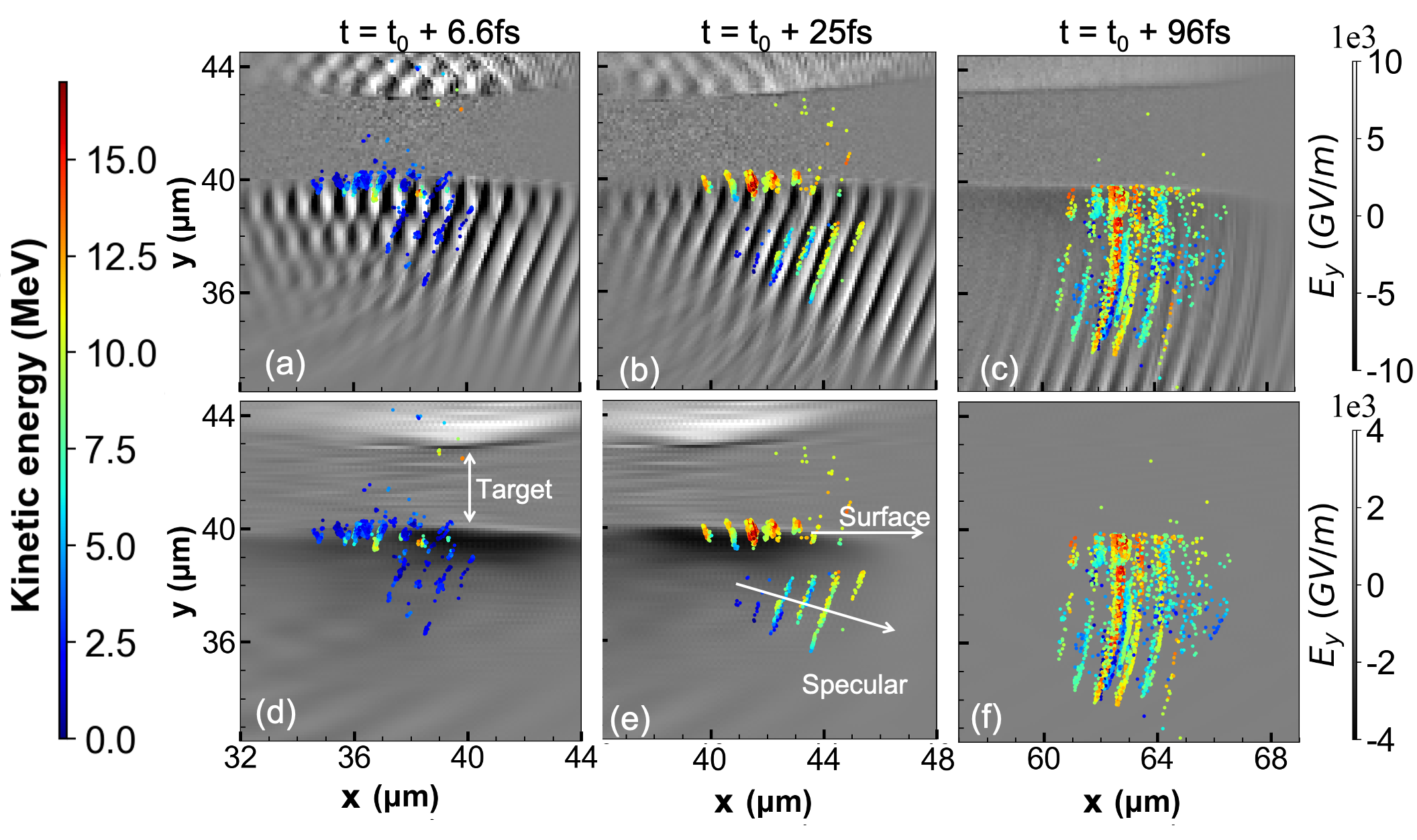}
   \caption{The top row illustrates the instantaneous positions and energies of randomly selected particles overlaid on the combined electric field from both laser pulses. The bottom row shows the same particles overlaid on the static electric field, highlighting their spatial distribution and the development of the sheath structure as the particles undergo acceleration.}
   \label{fig:Tracks}
\end{figure*}

\par To demonstrate this, we simulated a set-up shown by the schematic diagram given in Fig.~\ref{fig:setup}. Here, we employ a two-laser system in which one laser (Sheath laser) is used to generate the sheath field, while the other (Driver laser) interacts with plasma in this field to accelerate electrons along the target surface. Thus, the rear-side quasi-static sheath fields, produced by interaction of first laser with the target front surface, mediates the effective electron acceleration when the second laser is incident on the target rear surface. The numerical simulations of this set-up are performed by a 2D3V PIC code SMILEI\cite{derouillat2018smilei} on the high-performance computing (HPC) cluster facility at TIFR, Mumbai. The simulations are performed in two space (2D) dimensions as 3D simulations are extremely computationally expensive. Moreover, previous studies \cite{macchi2013ion} have shown that 2D3V simulations are sufficient to capture the essential physics of sheath formation and electron acceleration at the back of a target.
\par The lasers interact with a $100n_c$ solid-density plasma, where the critical density $n_c$\cite{Gibbon_1996} is given by:
$n_c [\text{cm}^{-3}] = 1.1 \times 10^{21} \left(\frac{1 }{\lambda_{L} (\mu \text{m})}\right)^2$ as shown in Fig.~\ref{fig:setup}.
The top laser ($I_{sheath}$) is incident at $45^\circ$ to maximize energy coupling into the plasma \cite{Electron_acc} and generate a strong sheath at the rear surface of a $2\lambda$ target. A second laser ($I_{driver}$), temporally synchronized, strikes at $70^\circ$ \cite{grazingincidence,grazing_inc,PRL_ZMsheng} in the presence of this sheath. Both are P-polarized with $a_0=3$, focused to $5\lambda$, and have a pulse duration of 12 cycles, consistent with experimental conditions. Additional simulation details can be seen in the methods section at the end of the paper. To minimize dependence on laser contrast, a 200 nm preplasma scale length \cite{denavit1979collisionless,contrast_enh} is added to both sides of the target, which facilitates improved laser–matter coupling. A virtual screen, indicated by a dotted line, was placed at $38.4 \mu\text{m}$ to record the energy and spatial distribution of electrons throughout the simulation.
\par To gain deeper insight into the interaction dynamics, it is essential to characterize the sheath field. Experimental techniques such as particle probing of rear foils~\cite{particle_probing_sheath} have provided valuable information, though the short timescales make field evolution difficult to resolve. The sheath field can be approximated as~\cite{Borgeshi_sheath_formula,formula_og_paper} $E \sim \frac{k T_{\text{hot}}/e}{\lambda_{\text{pl}}}$,
where $T_{\text{hot}}$ is the rear-surface electron temperature and $\lambda_{\text{pl}}$ is the density scale length at the rear side of the target. For a laser intensity $I\lambda^{2} \sim 10^{19}~\text{W/cm}^2\mu\text{m}^2$ ($a_0 \sim 3$), taking $T_{\text{hot}}=1.5~\text{MeV}$ and $\lambda_{\text{ion}} \sim 2~\mu$m gives a sheath field of several~TV/m. Simulations confirm this estimate, showing a peak sheath field of $\sim0.2\,$, consistent with the analytical model~\cite{formula_og_paper}. Details of the field extraction are provided in the Supplementary Information.
\begin{figure}[!ht]
   \centering
   \includegraphics[width=\linewidth]{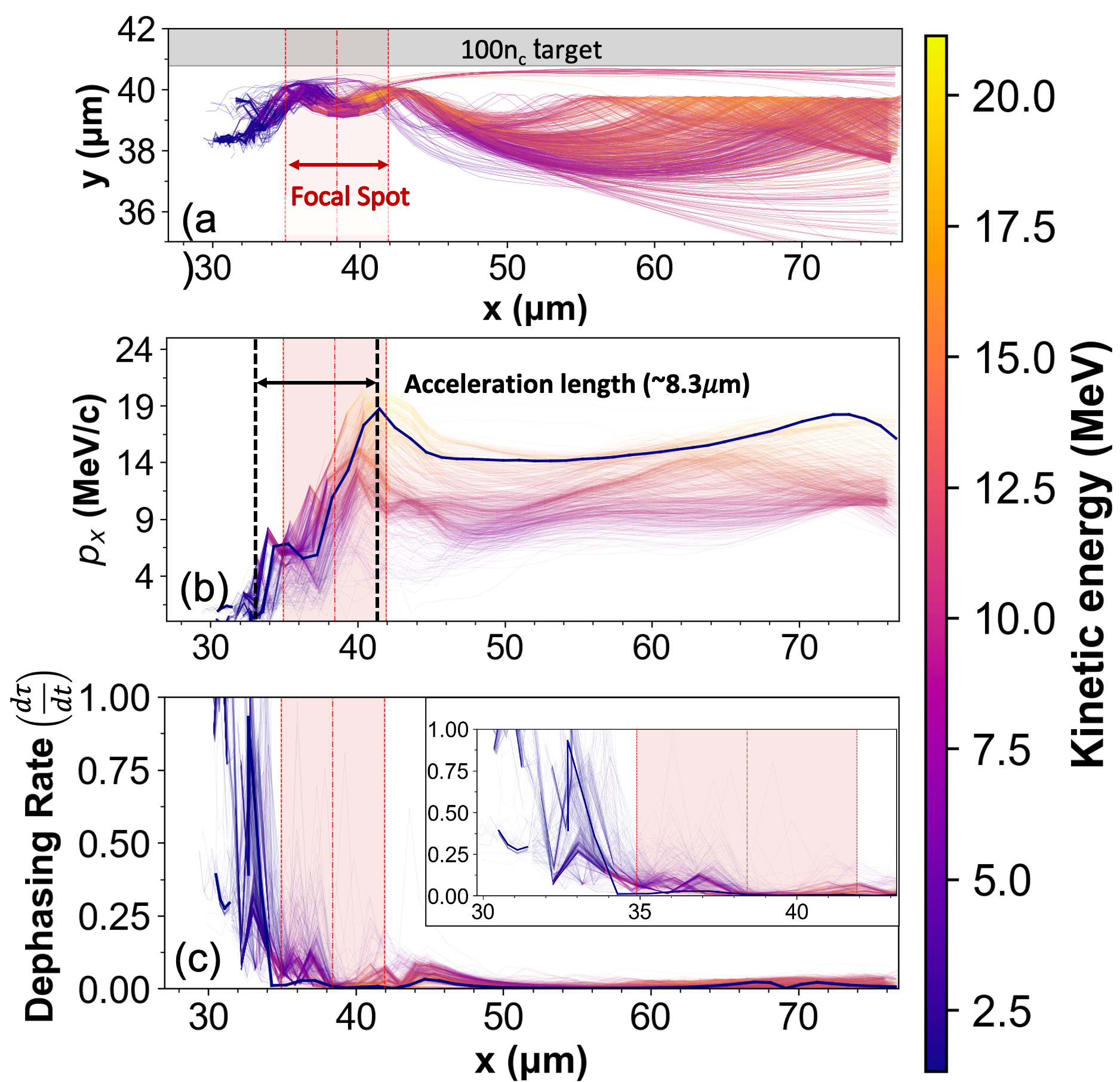}
   \caption{Panel (a) shows electron trajectories and panel (b) the corresponding phase space (c) the dephasing rate.}
   \label{fig:Sim_Spectrum}
\end{figure}

\begin{figure*}[!ht]
   \centering
   \includegraphics[width=\linewidth]{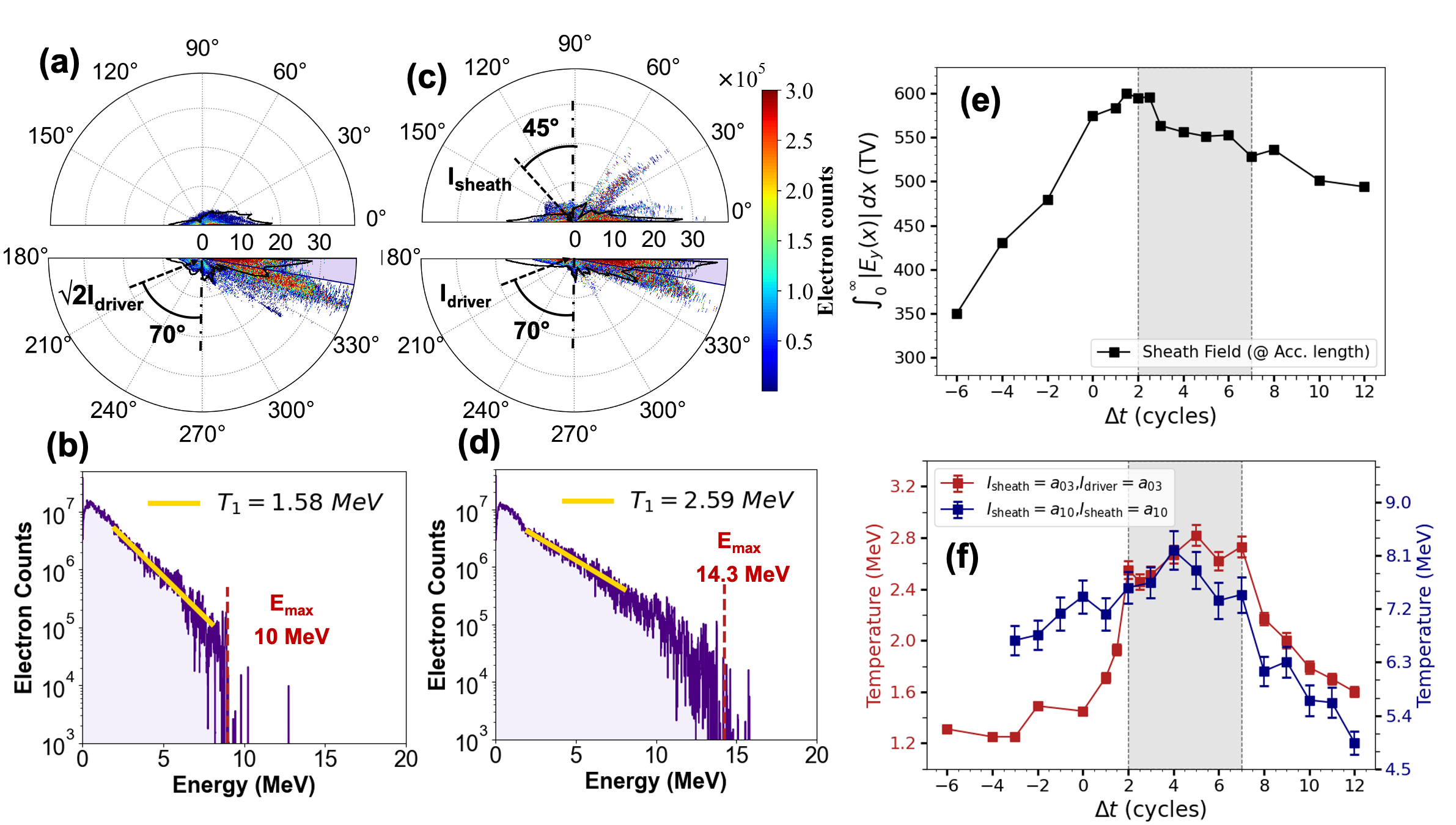}
   \caption{Panel (a): angular distribution of electrons for single-laser configuration. (c) Angular distribution for the dual-laser configuration. Panels (b,d) show the corresponding electron energy spectra extracted from a 10° cone around the surface normal on the positive-y side (shaded purple region in panels a,c). Panel (e) illustrates the evolution of the sheath field for a laser of $a_0=3$ as particles approach the acceleration region. Panel (f) displays the corresponding electron temperatures for both laser configurations: $a_0 = 3$ and $a_0 = 10$, highlighting the influence of laser intensity on particle acceleration.}
   \label{fig:ang_dist}
\end{figure*}
\par The sheath field is inherently dynamic, decaying in time and spreading laterally\cite{sheath_expansion}. However on the time-scale of short pulse laser-plasma interaction, it can be treated as quasi-static. This can be seen from Figure~\ref{fig:Tracks}, where we have plotted the evolution of $E_y$ at various time-steps. In particular, the bottom row shows the quasi-static fields after removing the oscillating laser electric field using the Fourier filtering. In addition, we have superimposed locations of $\sim$1,000 electrons with energies above 20~MeV. These electrons are color-coded as per their kinetic energies (shown by the left colorbar). In particular, Fig.~\ref{fig:Tracks}(e) reveals two distinct beams: one guided along the target surface, and another following the laser (specular) direction. Specifically, we see that the electron traveling along the target surface get phase-locked with the laser when the laser plasma interaction is strongest between $t_0 + 6.6$ fs to $t_0 + 25$ fs (see Fig.~\ref{fig:Tracks} (a) and (b)). The bunching of electrons seen in every half cycle (see Fig. \ref{fig:Tracks}) clearly demonstrates the phase locking of these electrons in presence of quasi static fields, consistent with the analytical predictions of equation \ref{eq:EOM_px}.

\par DLA-like acceleration of the electrons is demonstrated with Figure~\ref{fig:Sim_Spectrum}, where the acceleration dynamics of electrons is plotted. Panel (a)-(c) show the electron's spatial trajectories, the phase-space distribution along the acceleration direction, and the corresponding  dephasing rate ($d\tau/dt$),respectively. A sharp acceleration within a length of $\sim$8.3~$\mu$m, centered at the focal spot of the driver laser, confirms that electron acceleration is confined to a highly localized region. This contrasts with the single-laser case (see Supplementary Information), where particles alternately gain and lose energy without coherent synergy.

The transverse sheath field ($E_y$), while not directly depositing energy, plays a pivotal role by reducing the dephasing rate and phase-locking electrons to the laser, thereby sustaining efficient energy transfer. As shown in Fig.~\ref{fig:Tracks}(e), it drives an early split of electron bunches into surface-guided and specular components, enabling them to gain energy from the driving laser well before the main acceleration phase. This decisive mechanism—absent in the single-laser case—underpins the enhanced electron acceleration observed. This acceleration scheme is effective only when the sheath field lies in the plane of the laser polarization. The electrons generated by the S- polarized driving laser are shown in the  do not gain energy in the presence of the perpendicular plane transverse sheath field.

To support that, Figs.~\ref{fig:ang_dist}(a) show the angular distribution of electrons for single-laser setup with equal total incident energy. In contrast, Fig.~\ref{fig:ang_dist}(c) presents the two-laser configuration, where precise temporal synchronization leads to a clear enhancement of electron acceleration. In the angular plots, the radial coordinate represents the electron Lorentz factor ($\gamma$), while the black curve shows the total flux above 5~MeV integrated over all angles. The shaded purple region marks the $10^\circ$ sector used for further analysis. The corresponding energy spectra, integrated over this sector, are shown in the bottom row. Figure~\ref{fig:ang_dist}(b) is consistent with earlier results~\cite{Sarma_2022}, serving as a benchmark, whereas Fig.~\ref{fig:ang_dist}(d) demonstrates a clear increase in cutoff energy of $\sim$5--6~MeV with two synchronized pulses. The spectra are fitted with Maxwellian distributions to extract the hot electron temperature. A $64\%$ increase in temperature is observed compared to both single-laser cases, despite identical total laser energy on target.
In addition to angular distributions, energy coupling into electrons is enhanced by $\sim$34\% at $a_0=3$ (focal spot $3\lambda$, 12-cycle pulse, 30~mJ input)~\cite{YPing_abs,PRX_quere_abs}. The overall absorption into electrons rises from $\sim$24\% for the single-laser case to $\sim$27\% with two lasers (15~mJ each), clearly demonstrating the advantage of the dual-laser geometry.

 An important question is the sensitivity of electron heating to the timing between the two laser pulses. Figure~\ref{fig:ang_dist} shows the electron temperature, obtained from Maxwellian fits to the spectra, as a function of pulse delay. Unlike spectral cutoffs, which can be skewed by a few high-energy particles, temperature provides a more reliable measure of plasma heating. The red curve ($a_0=3$) in Fig.~\ref{fig:ang_dist} shows a twofold temperature enhancement for delays of $\sim$2–8 cycles (5.5–22~fs). The black curve ($a_0=10$) exhibits a similar window with smaller enhancement, since the higher-intensity leading edge already generates a sheath before zero delay. The onset of enhancement at $\sim$2 cycles reflects the finite time needed for sheath formation in a $2\lambda$-thick target, underscoring the critical role of timing in optimizing electron heating.

Our results establish that a transverse electric field, irrespective of its polarity, can profoundly reshape electron dynamics and, when appropriately tailored, provide a robust pathway for generating super-ponderomotive electrons. Building on this insight, we introduce a novel acceleration geometry that harnesses the sheath field—often considered detrimental to electron acceleration—as an effective accelerating structure. This approach is particularly advantageous for structured targets, in which scaling and precisely controlling the preplasma can present significant experimental and fabrication challenges. Compared with schemes that rely on costly fabrication, specialized targets or stringent laser parameters, the proposed method offers a simpler, more practical route to enhanced electron acceleration. These findings open new avenues for the flexible and scalable generation of energetic, directional electron beams using compact laser–plasma platforms.
 \vspace{0.5mm}
\\

\textbf{Methods}\\
The simulation was performed using normalized units, with the laser wavelength $\lambda = 800\,\mathrm{nm}$ as the fundamental length scale. A two-laser configuration was implemented using the Particle-In-Cell (PIC) code \textsc{Smilei}. The simulation domain was defined as $128\lambda \times 96\lambda$, discretized into $8192 \times 6144 $ cells. The plasma target was initialized with a peak electron density of $100\,n_c$, using 32 particles per cell. Second-order interpolation was employed for particle weighting, and the initial electron temperature was set isotropically to $1\,\mathrm{eV}$. Both lasers were modeled as linearly P - polarized (Electric field in the plane of the simulation) Gaussian pulses with a full width at half maximum (FWHM) of 12 optical cycles and a total duration of 48 cycles. Each pulse was focused to a Gaussian spot size of $3\lambda$. The total simulation time was $180\,t_0$, where $t_0 = \lambda/c \approx 2.66\,\mathrm{fs}$, allowing sufficient time for accelerated electrons to propagate through the entire simulation box. To analyze the angular and energy-resolved distribution of emitted electrons, virtual hemispherical diagnostics were placed at a distance of $48\lambda$ from the interaction point, both above and below the target, mimicking full-space electron detection.\\

\textbf{Data Availability}\\
The datasets generated and/or analyzed during the current study are available from the corresponding author upon reasonable request.
\\

\textbf{Code Availability}\\
The Particle-In-Cell simulations were performed using the open-source code \textsc{Smilei}, which is publicly available at: https://github.com/SmileiPIC/Smilei. Custom test-particle analysis codes developed during this study are available from the corresponding author upon reasonable request.
\bibliographystyle{apsrev4-2}
\bibliography{Main_paper/Main}

@article{Gibbon_1996,
author = {P Gibbon and E Förster},
doi = {10.1088/0741-3335/38/6/001},
url = {https://dx.doi.org/10.1088/0741-3335/38/6/001},
year = {1996},
month = {jun},
publisher = {},
volume = {38},
number = {6},
pages = {769},
title = {Short-pulse laser - plasma interactions},
journal = {Plasma Physics and Controlled Fusion}
}

@article{Wilks1993,
  author  = {Wilks, S. C.},
  title   = {Simulations of Ultraintense Laser--Plasma Interactions},
  journal = {Physics of Fluids B: Plasma Physics},
  volume  = {5},
  number  = {7},
  pages   = {2603--2608},
  year    = {1993},
  month   = jul,
  doi     = {10.1063/1.860697},
  url     = {https://doi.org/10.1063/1.860697}
}

@article{wilks2001energetic,
  title={Energetic proton generation in ultra-intense laser--solid interactions},
  author={Wilks, SC and Langdon, AB and Cowan, TE and Roth, M and Singh, M and Hatchett, S and Key, MH and Pennington, D and MacKinnon, A and Snavely, RA},
  journal={Physics of plasmas},
  volume={8},
  number={2},
  pages={542--549},
  year={2001},
  publisher={American Institute of Physics}
}

@article{hatchett2000electron,
  title={Electron, photon, and ion beams from the relativistic interaction of Petawatt laser pulses with solid targets},
  author={Hatchett, Stephen P and Brown, Curtis G and Cowan, Thomas E and Henry, Eugene A and Johnson, Joy S and Key, Michael H and Koch, Jeffrey A and Langdon, A Bruce and Lasinski, Barbara F and Lee, Richard W and others},
  journal={Physics of Plasmas},
  volume={7},
  number={5},
  pages={2076--2082},
  year={2000},
  publisher={American Institute of Physics}
}

@article{snavely2000intense,
  title={Intense high-energy proton beams from petawatt-laser irradiation of solids},
  author={Snavely, RA and Key, MH and Hatchett, SP and Cowan, TE and Roth, Markus and Phillips, TW and Stoyer, MA and Henry, EA and Sangster, TC and Singh, MS and others},
  journal={Physical review letters},
  volume={85},
  number={14},
  pages={2945},
  year={2000},
  publisher={APS}
}

@article{denavit1979collisionless,
  title={Collisionless plasma expansion into a vacuum},
  author={Denavit, J},
  journal={The Physics of Fluids},
  volume={22},
  number={7},
  pages={1384--1392},
  year={1979},
  publisher={American Institute of Physics}
}

@article{derouillat2018smilei,
  title={Smilei: A collaborative, open-source, multi-purpose particle-in-cell code for plasma simulation},
  author={Derouillat, Julien and Beck, Arnaud and P{\'e}rez, Fr{\'e}d{\'e}ric and Vinci, Tommaso and Chiaramello, M and Grassi, Anna and Fl{\'e}, M and Bouchard, Guillaume and Plotnikov, I and Aunai, Nicolas and others},
  journal={Computer Physics Communications},
  volume={222},
  pages={351--373},
  year={2018},
  publisher={Elsevier}
}

@article{Macchi,
author = {Macchi,A. },
title = {Surface plasmons in superintense laser-solid interactions},
journal = {Physics of Plasmas},
volume = {25},
number = {3},
pages = {031906},
year = {2018},
doi = {10.1063/1.5013321},
URL = {https://doi.org/10.1063/1.5013321},
eprint = {https://doi.org/10.1063/1.5013321}
}

@article{Riconda_PoP2015_SPW_electron_scaling,

author = {Riconda, C. and Raynaud, M. and Vialis, T. and Grech, M.},
title = "{Simple scalings for various regimes of electron acceleration in surface plasma waves}",
journal = {Physics of Plasmas},
volume = {22},
number = {7},
pages = {073103},
year = {2015},
month = {07},
issn = {1070-664X},
doi = {10.1063/1.4923443},
url = {https://doi.org/10.1063/1.4923443},
}

@article{nilson2008high,
  title={High-intensity laser-plasma interactions in the refluxing limit},
  author={Nilson, PM and Theobald, W and Myatt, J and Stoeckl, C and Storm, M and Gotchev, OV and Zuegel, JD and Betti, R and Meyerhofer, DD and Sangster, TC},
  journal={Physics of Plasmas},
  volume={15},
  number={5},
  year={2008},
  publisher={AIP Publishing}
}

@article{buffechoux2010hot,
  title={Hot electrons transverse refluxing in ultraintense laser-solid interactions},
  author={Buffechoux, S and Psikal, J and Nakatsutsumi, M and Romagnani, L and Andreev, A and Zeil, K and Amin, M and Antici, P and Burris-Mog, T and Compant-La-Fontaine, A and others},
  journal={Physical review letters},
  volume={105},
  number={1},
  pages={015005},
  year={2010},
  publisher={APS}
}

@Article{Malka2008,
author={Malka, Victor
and Faure, J{\'e}r{\^o}me
and Gauduel, Yann A.
and Lefebvre, Erik
and Rousse, Antoine
and Phuoc, Kim Ta},
title={Principles and applications of compact laser--plasma accelerators},
journal={Nature Physics},
year={2008},
month={Jun},
day={01},
volume={4},
number={6},
pages={447-453},
issn={1745-2481},
doi={10.1038/nphys966},
url={https://doi.org/10.1038/nphys966}
}

@article{Electron_acc,
  title = {Experimental Confirmation of Ponderomotive-Force Electrons Produced by an Ultrarelativistic Laser Pulse on a Solid Target},
  author = {Malka, G. and Miquel, J. L.},
  journal = {Phys. Rev. Lett.},
  volume = {77},
  issue = {1},
  pages = {75--78},
  numpages = {0},
  year = {1996},
  month = {Jul},
  publisher = {American Physical Society},
  doi = {10.1103/PhysRevLett.77.75},
  url = {https://link.aps.org/doi/10.1103/PhysRevLett.77.75}
}

@article{macchi2013ion,
  title={Ion acceleration by superintense laser-plasma interaction},
  author={Macchi, Andrea and Borghesi, Marco and Passoni, Matteo},
  journal={Reviews of Modern Physics},
  volume={85},
  number={2},
  pages={751--793},
  year={2013},
  publisher={APS}
}

@InProceedings{Borgeshi_sheath_formula,
author="Borghesi, Marco",
editor="Gizzi, Leonida Antonio
and Assmann, Ralph
and Koester, Petra 
and Giulietti, Antonio",
title="Ion Acceleration: TNSA and Beyond",
booktitle="Laser-Driven Sources of High Energy Particles and Radiation",
year="2019",
publisher="Springer International Publishing",
address="Cham",
pages="143--164",
isbn="978-3-030-25850-4"
}

@article{formula_og_paper,
  title = {Forward Ion Acceleration in Thin Films Driven by a High-Intensity Laser},
  author = {Maksimchuk, A. and Gu, S. and Flippo, K. and Umstadter, D. and Bychenkov, V. Yu.},
  journal = {Phys. Rev. Lett.},
  volume = {84},
  issue = {18},
  pages = {4108--4111},
  numpages = {0},
  year = {2000},
  month = {May},
  publisher = {American Physical Society},
  doi = {10.1103/PhysRevLett.84.4108},
  url = {https://link.aps.org/doi/10.1103/PhysRevLett.84.4108}
}

@article{particle_probing_sheath,
  title = {Dynamics of Electric Fields Driving the Laser Acceleration of Multi-MeV Protons},
  author = {Romagnani, L. and Fuchs, J. and Borghesi, M. and Antici, P. and Audebert, P. and Ceccherini, F. and Cowan, T. and Grismayer, T. and Kar, S. and Macchi, A. and Mora, P. and Pretzler, G. and Schiavi, A. and Toncian, T. and Willi, O.},
  journal = {Phys. Rev. Lett.},
  volume = {95},
  issue = {19},
  pages = {195001},
  numpages = {4},
  year = {2005},
  month = {Oct},
  publisher = {American Physical Society},
  doi = {10.1103/PhysRevLett.95.195001},
  url = {https://link.aps.org/doi/10.1103/PhysRevLett.95.195001}
}

@article{PRX_quere_abs,
  title = {Identification of Coupling Mechanisms between Ultraintense Laser Light and Dense Plasmas},
  author = {Chopineau, L. and Leblanc, A. and Blaclard, G. and Denoeud, A. and Th\'evenet, M. and Vay, J-L. and Bonnaud, G. and Martin, Ph. and Vincenti, H. and Qu\'er\'e, F.},
  journal = {Phys. Rev. X},
  volume = {9},
  issue = {1},
  pages = {011050},
  numpages = {18},
  year = {2019},
  month = {Mar},
  publisher = {American Physical Society},
  doi = {10.1103/PhysRevX.9.011050},
  url = {https://link.aps.org/doi/10.1103/PhysRevX.9.011050}
}

@article{YPing_abs,
  title = {Absorption of Short Laser Pulses on Solid Targets in the Ultrarelativistic Regime},
  author = {Ping, Y. and Shepherd, R. and Lasinski, B. F. and Tabak, M. and Chen, H. and Chung, H. K. and Fournier, K. B. and Hansen, S. B. and Kemp, A. and Liedahl, D. A. and Widmann, K. and Wilks, S. C. and Rozmus, W. and Sherlock, M.},
  journal = {Phys. Rev. Lett.},
  volume = {100},
  issue = {8},
  pages = {085004},
  numpages = {4},
  year = {2008},
  month = {Feb},
  publisher = {American Physical Society},
  doi = {10.1103/PhysRevLett.100.085004},
  url = {https://link.aps.org/doi/10.1103/PhysRevLett.100.085004}
}

@article{VLA_natphy,
  title={Vacuum laser acceleration of relativistic electrons using plasma mirror injectors},
  author={Th{\'e}venet, M and Leblanc, A and Kahaly, S and Vincenti, H and Vernier, A and Qu{\'e}r{\'e}, F and Faure, J{\'e}r{\^o}me},
  journal={Nature Physics},
  volume={12},
  number={4},
  pages={355--360},
  year={2016},
  publisher={Nature Publishing Group UK London}
}

@article{VLA_OG,
  title = {Laser acceleration of electrons in vacuum},
  author = {Esarey, Eric and Sprangle, Phillip and Krall, Jonathan},
  journal = {Phys. Rev. E},
  volume = {52},
  issue = {5},
  pages = {5443--5453},
  numpages = {0},
  year = {1995},
  month = {Nov},
  publisher = {American Physical Society},
  doi = {10.1103/PhysRevE.52.5443},
  url = {https://link.aps.org/doi/10.1103/PhysRevE.52.5443}
}

@article{VLA_semiinfinite,
  title = {Visible-Laser Acceleration of Relativistic Electrons in a Semi-Infinite Vacuum},
  author = {Plettner, T. and Byer, R. L. and Colby, E. and Cowan, B. and Sears, C. M. S. and Spencer, J. E. and Siemann, R. H.},
  journal = {Phys. Rev. Lett.},
  volume = {95},
  issue = {13},
  pages = {134801},
  numpages = {4},
  year = {2005},
  month = {Sep},
  publisher = {American Physical Society},
  doi = {10.1103/PhysRevLett.95.134801},
  url = {https://link.aps.org/doi/10.1103/PhysRevLett.95.134801}
}

@article{VLAsingh2022vacuum,
  title={Vacuum laser acceleration of super-ponderomotive electrons using relativistic transparency injection},
  author={Singh, Prashant Kumar and Li, F-Y and Huang, C-K and Moreau, Adam and Hollinger, Reed and Junghans, A and Favalli, Andrea and Calvi, C and Wang, Shoujun and Wang, Y and others},
  journal={Nature Communications},
  volume={13},
  number={1},
  pages={54},
  year={2022},
  publisher={Nature Publishing Group UK London}
}

@article{Two_laser_electron,
  title = {Highly intensified emission of laser-accelerated electrons from a foil target through an additional rear laser plasma},
  author = {Inoue, Shunsuke and Nakamiya, Yoshihide and Teramoto, Kensuke and Hashida, Masaki and Sakabe, Shuji},
  journal = {Phys. Rev. Accel. Beams},
  volume = {21},
  issue = {4},
  pages = {041302},
  numpages = {6},
  year = {2018},
  month = {Apr},
  publisher = {American Physical Society},
  doi = {10.1103/PhysRevAccelBeams.21.041302},
  url = {https://link.aps.org/doi/10.1103/PhysRevAccelBeams.21.041302}
}

@article{grazingincidence,
    author = {Serebryakov, D. A. and Nerush, E. N. and Kostyukov, I. Yu.},
    title = {Near-surface electron acceleration during intense laser–solid interaction in the grazing incidence regime},
    journal = {Physics of Plasmas},
    volume = {24},
    number = {12},
    pages = {123115},
    year = {2017},
    month = {12},
    issn = {1070-664X},
    doi = {10.1063/1.5002671},
    url = {https://doi.org/10.1063/1.5002671},
    eprint = {https://pubs.aip.org/aip/pop/article-pdf/doi/10.1063/1.5002671/13307528/123115\_1\_online.pdf},
}

@article{grazing_inc,
    author = {Yabuuchi, T. and Adumi, K. and Habara, H. and Kodama, R. and Kondo, K. and Tanimoto, T. and Tanaka, K. A. and Sentoku, Y. and Matsuoka, T. and Chen, Z. L. and Tampo, M. and Lei, A. L. and Mima, K.},
    title = {On the behavior of ultraintense laser produced hot electrons in self-excited fields},
    journal = {Physics of Plasmas},
    volume = {14},
    number = {4},
    pages = {040706},
    year = {2007},
    month = {04},
    issn = {1070-664X},
    doi = {10.1063/1.2722303},
    url = {https://doi.org/10.1063/1.2722303},
    eprint = {https://pubs.aip.org/aip/pop/article-pdf/doi/10.1063/1.2722303/14867060/040706\_1\_online.pdf},
}

@article{Sarma_2022,
doi = {10.1088/1367-2630/ac7d6e},
url = {https://dx.doi.org/10.1088/1367-2630/ac7d6e},
year = {2022},
month = {jul},
publisher = {IOP Publishing},
volume = {24},
number = {7},
pages = {073023},
author = {Sarma, J and McIlvenny, A and Das, N and Borghesi, M and Macchi, A},
title = {Surface plasmon-driven electron and proton acceleration without grating coupling},
journal = {New Journal of Physics}
}

@article{woodwork,
author = {P.M. Woodward },
title = {A method of calculating the field over a plane aperture required to produce a given polar diagram},
journal = {Journal of the Institution of Electrical Engineers - Part IIIA: Radiolocation},
volume = {93},
issue = {10},
pages = {1554-1558},
year = {1946},
doi = {10.1049/ji-3a-1.1946.0262},

URL = {https://digital-library.theiet.org/doi/abs/10.1049/ji-3a-1.1946.0262},
eprint = {https://digital-library.theiet.org/doi/pdf/10.1049/ji-3a-1.1946.0262}
}

@ARTICLE{Lawson,
  author={Lawson, J D},
  journal={IEEE Transactions on Nuclear Science}, 
  title={Lasers and Accelerators}, 
  year={1979},
  volume={26},
  number={3},
  pages={4217-4219},
  doi={10.1109/TNS.1979.4330749}}

@article{ext_mag_1,
  title = {Energy gain by laser-accelerated electrons in a strong magnetic field},
  author = {Arefiev, A. and Gong, Z. and Robinson, A. P. L.},
  journal = {Phys. Rev. E},
  volume = {101},
  issue = {4},
  pages = {043201},
  numpages = {10},
  year = {2020},
  month = {Apr},
  publisher = {American Physical Society},
  doi = {10.1103/PhysRevE.101.043201},
  url = {https://link.aps.org/doi/10.1103/PhysRevE.101.043201}
}

@article{ext_mag_2,
  author  = {Robinson, A. P. L. and Arefiev, A. V.},
  title   = {Net Energy Gain in Direct Laser Acceleration Due to Enhanced Dephasing Induced by an Applied Magnetic Field},
  journal = {Physics of Plasmas},
  volume  = {27},
  number  = {2},
  pages   = {023110},
  year    = {2020},
  month   = feb,
  doi     = {10.1063/1.5122893},
  url     = {https://doi.org/10.1063/1.5122893}
}

@article{ext_mag_3_axial,
  author  = {Yang, X. H. and Borghesi, M. and Qiao, B. and Geissler, M. and Robinson, A. P. L.},
  title   = {Effects of External Axial Magnetic Field on Fast Electron Propagation},
  journal = {Physics of Plasmas},
  volume  = {18},
  number  = {9},
  pages   = {093102},
  year    = {2011},
  month   = sep,
  doi     = {10.1063/1.3630925},
  url     = {https://doi.org/10.1063/1.3630925}
}

@article{nature_refluxing_exp,
	author = {Green, J. S. and Booth, N. and Dance, R. J. and Gray, R. J. and MacLellan, D. A. and Marshall, A. and McKenna, P. and Murphy, C. D. and Ridgers, C. P. and Robinson, A. P. L. and Rusby, D. and Scott, R. H. H. and Wilson, L.},
	date = {2018/03/14},
	doi = {10.1038/s41598-018-22422-6},
	id = {Green2018},
	isbn = {2045-2322},
	journal = {Scientific Reports},
	number = {1},
	pages = {4525},
	title = {Time-resolved measurements of fast electron recirculation for relativistically intense femtosecond scale laser-plasma interactions},
	url = {https://doi.org/10.1038/s41598-018-22422-6},
	volume = {8},
	year = {2018}}

@article{ext_long_1,
  title = {Generating ``Superponderomotive'' Electrons due to a Non-Wake-Field Interaction between a Laser Pulse and a Longitudinal Electric Field},
  author = {Robinson, A. P. L. and Arefiev, A. V. and Neely, D.},
  journal = {Phys. Rev. Lett.},
  volume = {111},
  issue = {6},
  pages = {065002},
  numpages = {5},
  year = {2013},
  month = {Aug},
  publisher = {American Physical Society},
  doi = {10.1103/PhysRevLett.111.065002},
  url = {https://link.aps.org/doi/10.1103/PhysRevLett.111.065002}
}

@article{ext_long_2,
    author = {Arefiev, A. V. and Khudik, V. N. and Robinson, A. P. L. and Shvets, G. and Willingale, L. and Schollmeier, M.},
    title = {Beyond the ponderomotive limit: Direct laser acceleration of relativistic electrons in sub-critical plasmas},
    journal = {Physics of Plasmas},
    volume = {23},
    number = {5},
    pages = {056704},
    year = {2016},
    month = {04},
    issn = {1070-664X},
    doi = {10.1063/1.4946024},
    url = {https://doi.org/10.1063/1.4946024},
    eprint = {https://pubs.aip.org/aip/pop/article-pdf/doi/10.1063/1.4946024/15946709/056704\_1\_online.pdf},
}

@article{sheath_expansion,
  title = {Superluminal sheath-field expansion and fast-electron-beam divergence measurements in laser-solid interactions},
  author = {Ridgers, C. P. and Sherlock, M. and Evans, R. G. and Robinson, A. P. L. and Kingham, R. J.},
  journal = {Phys. Rev. E},
  volume = {83},
  issue = {3},
  pages = {036404},
  numpages = {10},
  year = {2011},
  month = {Mar},
  publisher = {American Physical Society},
  doi = {10.1103/PhysRevE.83.036404},
  url = {https://link.aps.org/doi/10.1103/PhysRevE.83.036404}
}

@article{Optica_rocca,
author = {Jorge J. Rocca and Maria G. Capeluto and Reed C. Hollinger and Shoujun Wang and Yong Wang and G. Ravindra Kumar and Amit D. Lad and Alexander Pukhov and Vyacheslav N. Shlyaptsev},
journal = {Optica},
number = {3},
pages = {437--453},
publisher = {Optica Publishing Group},
title = {Ultra-intense femtosecond laser interactions with aligned nanostructures},
volume = {11},
month = {Mar},
year = {2024},
url = {https://opg.optica.org/optica/abstract.cfm?URI=optica-11-3-437},
doi = {10.1364/OPTICA.510542},
}

@article{PRL_ZMsheng,
  title = {Observation of a Fast Electron Beam Emitted along the Surface of a Target Irradiated by Intense Femtosecond Laser Pulses},
  author = {Li, Y. T. and Yuan, X. H. and Xu, M. H. and Zheng, Z. Y. and Sheng, Z. M. and Chen, M. and Ma, Y. Y. and Liang, W. X. and Yu, Q. Z. and Zhang, Y. and Liu, F. and Wang, Z. H. and Wei, Z. Y. and Zhao, W. and Jin, Z. and Zhang, J.},
  journal = {Phys. Rev. Lett.},
  volume = {96},
  issue = {16},
  pages = {165003},
  numpages = {4},
  year = {2006},
  month = {Apr},
  publisher = {American Physical Society},
  doi = {10.1103/PhysRevLett.96.165003},
  url = {https://link.aps.org/doi/10.1103/PhysRevLett.96.165003}
}

@article{contrast_enh,
  title = {Spectrally peaked electron beams produced via surface guiding and acceleration in femtosecond laser-solid interactions},
  author = {Mao, J. Y. and Chen, L. M. and Ge, X. L. and Zhang, L. and Yan, W. C. and Li, D. Z. and Liao, G. Q. and Ma, J. L. and Huang, K. and Li, Y. T. and Lu, X. and Dong, Q. L. and Wei, Z. Y. and Sheng, Z. M. and Zhang, J.},
  journal = {Phys. Rev. E},
  volume = {85},
  issue = {2},
  pages = {025401},
  numpages = {4},
  year = {2012},
  month = {Feb},
  publisher = {American Physical Society},
  doi = {10.1103/PhysRevE.85.025401},
  url = {https://link.aps.org/doi/10.1103/PhysRevE.85.025401}
}

@article{tajima1979laser,
  title={Laser electron accelerator},
  author={Tajima, Toshiki and Dawson, John M},
  journal={Physical review letters},
  volume={43},
  number={4},
  pages={267},
  year={1979},
  publisher={APS}
}

@article{esarey2009physics,
  title={Physics of laser-driven plasma-based electron accelerators},
  author={Esarey, Eric and Schroeder, Carl B and Leemans, Wim P},
  journal={Reviews of modern physics},
  volume={81},
  number={3},
  pages={1229--1285},
  year={2009},
  publisher={APS}
}

@article{arefiev_paramatric_amplification,
  title = {Parametric Amplification of Laser-Driven Electron Acceleration in Underdense Plasma},
  author = {Arefiev, Alexey V. and Breizman, Boris N. and Schollmeier, Marius and Khudik, Vladimir N.},
  journal = {Phys. Rev. Lett.},
  volume = {108},
  issue = {14},
  pages = {145004},
  numpages = {5},
  year = {2012},
  month = {Apr},
  publisher = {American Physical Society},
  doi = {10.1103/PhysRevLett.108.145004},
  url = {https://link.aps.org/doi/10.1103/PhysRevLett.108.145004}
}

\clearpage
\onecolumngrid

\begin{center}
    {\Large\bfseries Supplementary - Quasi-static transverse electric field driven electron acceleration in relativistic laser matter interaction}
\end{center}
\section{Fourier Filtering}
To isolate the sheath field, Fourier filtering is applied since its intensity is several orders of magnitude lower than that of the incident and reflected laser fields. Figure~\ref{fig:Fourier}(a) shows the interaction phase containing both laser components, while the Fourier map in Fig.~\ref{fig:Fourier}(b) reveals distinct high-frequency harmonics. Removing these components exposes the quasi-static sheath field, as seen in the inverse transform [Fig.~\ref{fig:Fourier}(d)], with an amplitude of about 0.2 times the laser field.
\begin{figure}[h!]
   \centering
   \includegraphics[width=0.7\linewidth]{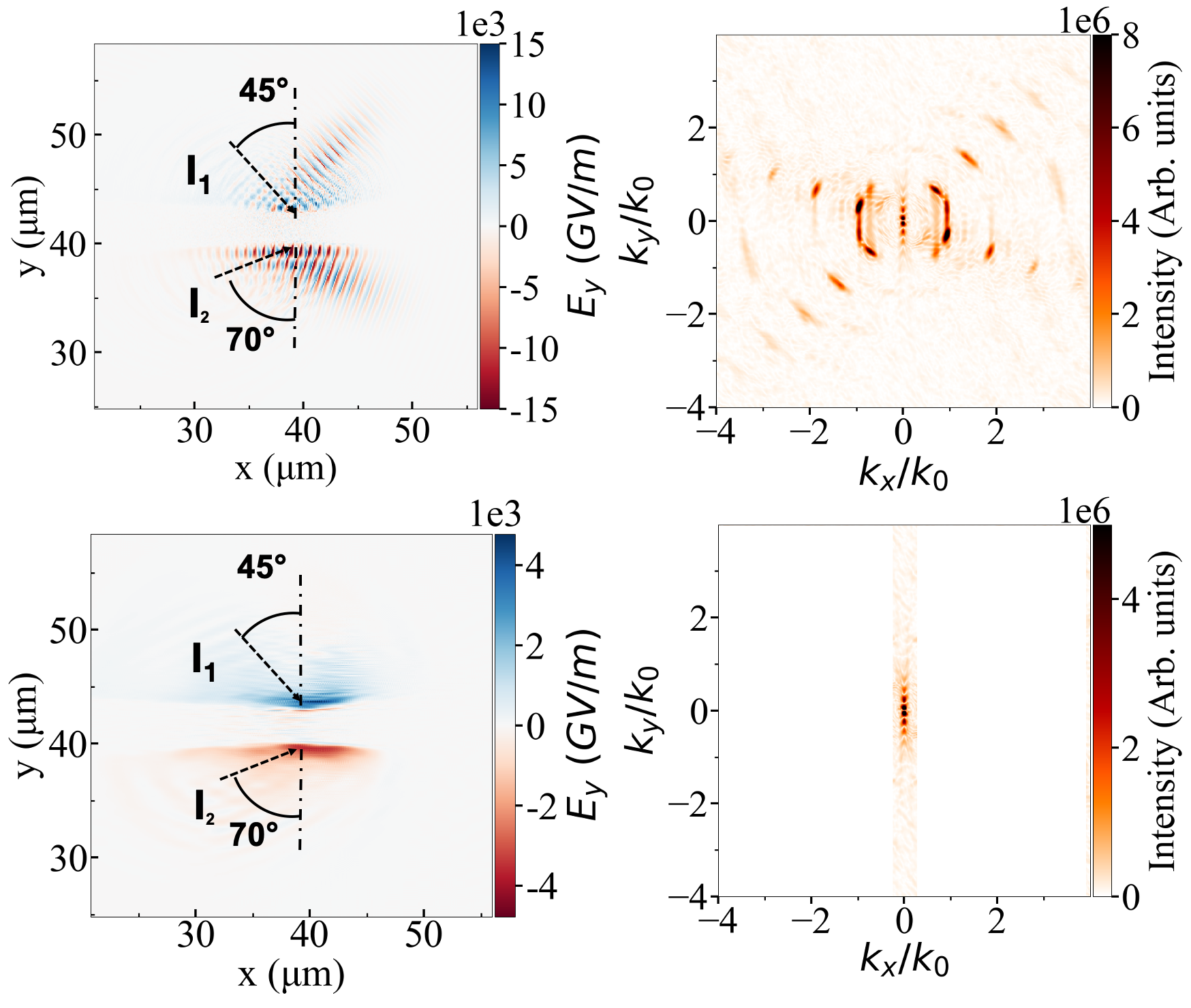}
    \captionsetup{justification=centering}
   \caption{Illustration of the Fourier filtering technique used to eliminate high-frequency oscillations from the laser field, enabling clear visualization of the underlying static field. This is achieved by applying a Fourier transform to the field data and removing frequency components deviating from zero in Fourier space. }
   \label{fig:Fourier}
\end{figure}

\section{Electron angular distribution using S polarized driving laser}
\begin{figure}[h!]
   \centering
   \includegraphics[width=0.7\linewidth]{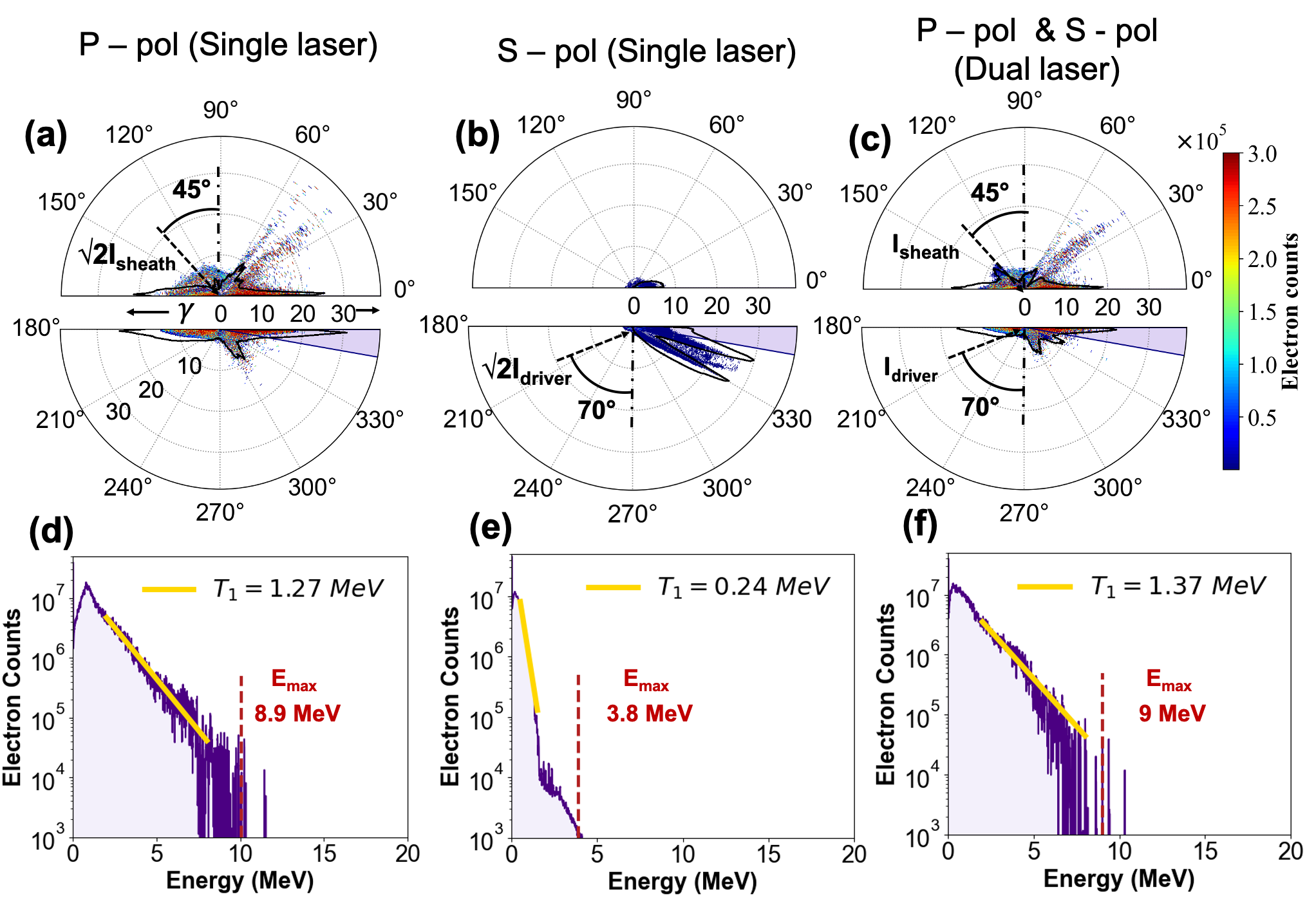}
    \captionsetup{justification=centering}
   \caption{ 2: (a, b)Angular distribution of electrons for single-laser configurations of P polarized (Electric field in the plane of simulation) and S polarized respectively. (c) Angular distribution for the dual-laser configuration. Panels (d, e, f) show the corresponding electron energy spectra extracted from a 10° cone around the surface normal on the positive-y side (shaded purple region in panels a–c). }
   \label{fig:Supp5_Spol}
\end{figure}
This section highlights the importance of polarization for the effectiveness of the acceleration scheme.  
Figure~\ref{fig:Supp5_Spol}(b) shows that, in the case of a single $S$-polarized laser, insufficient electrons are produced at the surface.  
Moreover, the electron quiver motion and the sheath field are oriented perpendicular to one another, which prevents efficient acceleration of the electrons. The flux and cutoff observed in Fig.~\ref{fig:Supp5_Spol}(c) remain essentially the same as in Fig.~\ref{fig:Supp5_Spol}(d), indicating that no additional mechanisms are contributing to the acceleration process.

\end{document}